\documentclass[a4paper,fleqn,usenatbib]{mnras}

\usepackage{newtxtext,newtxmath}

\usepackage[T1]{fontenc}
\usepackage{ae,aecompl}

\usepackage{graphicx}	
\usepackage{amsmath}	
\usepackage{amssymb}	

\usepackage{rotating}
\usepackage{threeparttable}

\usepackage{dcolumn}
\usepackage{longtable}
\usepackage{multirow}
\usepackage{bm}
\usepackage{textpos}

\usepackage{pdflscape} 

\title[An Empirical Fit for Viscoelastic Simulations of Tertiary Tides]{An Empirical Fit for Viscoelastic Simulations of Tertiary Tides}
\author[Y. Gao et al.]{
Yan Gao,$^{1,2,3}$\thanks{E-mail: ygbcyy@ynao.ac.cn}
Silvia Toonen,$^{4,5}$
Evgeni Grishin,$^{6}$
Tom Comerford,$^{7}$
Matthias U. Kruckow,$^{1,2}$
\\
$^{1}$Yunnan Observatories, Chinese Academy of Sciences, Kunming 650011, China\\
$^{2}$Key Laboratory for the Structure and Evolution of Celestial Objects, Chinese Academy of Sciences, Kunming 650011, China\\
$^{3}$University of Chinese Academy of Sciences\\
$^{4}$Anton Pannekoek Institute for Astronomy, University Of Amsterdam, 1090 GE Amsterdam, The Netherlands\\
$^{5}$Institute for Gravitational Wave Astronomy, School of Physics and Astronomy, University of Birmingham, B15 2TT Birmingham, UK\\
$^{6}$Physics Department, Technion - Israel Institute of Technology, Haifa 3200003, Israel\\
$^{7}$Institute of Astronomy, University of Camridge, Madingley Road, Cambridge CB3 OHA, UK\\
}

\date{Accepted XXX. Received YYY; in original form ZZZ}

\pubyear{2019}

\begin{document}
\label{firstpage}
\pagerange{\pageref{firstpage}--\pageref{lastpage}}
\maketitle

\begin{abstract}
Tertiary tides (TTs), or the continuous tidal distortion of the tertiary in a hierarchical triple system, can extract energy from the inner binary, inducing within it a proclivity to merge. Despite previous work on the subject, which established that it is significant for certain close triple systems, it is still not a well-understood process. A portion of our ignorance in this regard stems from our inability to integrate a simulation of this phenomenon into conventional stellar evolution codes, since full calculations of these tidal interactions are computationally expensive on stellar evolution timescales. Thus, to attain a better understanding of how these TTs act on longer timescales, an empirical expression of its effects as a function of parameters of the triple system involved is required. In our work, we evaluate the rate at which TTs extract energy from the inner binary within a series of constructed hierarchical triple systems under varying parameters, and study the rate at which the inner binary orbital separation shrinks as a function of those parameters. We find that this rate varies little with the absolute values of the masses of the three component objects, but is very sensitive to the mass ratio of the inner binary $q$, the tertiary radius $R_{\rm 3}$, the inner binary orbital separation $a_{\rm 1}$, the outer orbital separation $a_{\rm 2}$, and the viscoelastic relaxation time of the tertiary $\tau$. More specifically, we find that the percentage by which $a_{\rm 1}$ shrinks per unit time can be reasonably approximated by (1/$a_{\rm 1}$)(d$a_{\rm 1}$/d$t$)=$\left(2.22{\times}10^{-8}{\rm yrs}^{-1}\right)4q\left(1+q\right)^{-2}(R_{\rm 3}/100{\rm R}_{\odot})^{5.2}(a_{\rm 1}/0.2{\rm AU})^{4.8}(a_{\rm 2}/2{\rm AU})^{-10.2}$ $({\tau}/0.534{\rm yrs})^{-1.0}$. We also provide tests of how precise this fitting function is.
\end{abstract}




\section{Introduction}

Previous studies (\citealt{2018MNRAS.479.3604G}, henceforth GCEH18, see also \citealt{2013MNRAS.429.2425F}) have established the fact that, in a hierarchical triple system where the tertiary is sufficiently close to the inner binary, it extracts orbital energy from the inner binary via purely tidal interactions, giving rise within the inner binary a proclivity to merge. These tidal interactions are known as tertiary tides, or TTs for short. Since this merging process and its subsequent influence on the evolution of the triple system happens on timescales comparable to typical stellar evolution timescales, it would be desirable to integrate this process into a stellar evolution code. However, as the tidal effects leading to this merging process occur on timescales much shorter than stellar evolution timescales, it would be impractical to run a simulation of these tidal effects in parallel with a stellar evolution code. This complication is further compounded by the fact that triple evolution codes already need to deal with a host of complicated processes not seen in binaries, more of which are being discovered every year \citep[e.g.][]{2019AAS...23341405D}. The conventional way around such a problem would be to perform a set of simulations for a grid of parameters, so that detailed calculations can be replaced by interpolations and/or extrapolations of the aforementioned grid \citep[e.g.][]{2016ApJS..222....8D,2002MNRAS.329..897H,2000MNRAS.315..543H}, which can then be implemented in stellar evolution codes with relative ease. But since no such grid has yet been simulated for TTs, doing so would seem to be a natural course of action, hence this paper.

Since, in our previous studies, we found that the main influence of TTs is to quickly shrink the inner orbit of its host hierarchical triple, the speed at which this happens will be the main focus of the present work. Ideally, we wish to obtain ${\rm d}a_{\rm 1}/{\rm d}t$ as a function of the orbital parameters of the hierarchical triple, so that any triple stellar evolution code \citep[e.g.][]{2017AAS...22932605T} can easily implement this during the course of integrating the relevant stellar evolution functions, with little extra expenditure in terms of computing time. This function is therefore what we will aim to calculate empirically using our grid of simulations. Other effects, such as the seemingly negligible excitation of eccentricities of the inner and outer orbits, which may or may not be important, are still not well understood, and in any case do not affect the host triple system as obviously as the orbital shrinkage, so we leave them to future studies.

In this paper, we use our previous models to calculate the energy extraction rate for triples with varying orbital parameters, and provide an empirical fit to the results, thereby establishing the desired empirical function. In \S 2 we present our models and methods for calculating this energy extraction rate for individual systems, in \S 3 we display our results and our empirical fits to the results, and finally we discuss the implications of our work in \S 4.

\section{Methods}

\subsection{Simulations of Tertiary Tides}

To simulate a close hierarchical triple system undergoing TTs, we refer to our previous methods used in GCEH18. In that paper, the simulation is carried out in two stages. In the first stage, we used a lagged equilibrium tidal model  \citep{1981A&A....99..126H,1998ApJ...499..853E,1998MNRAS.300..292K, 2016CeMDA.126..189C} to estimate the magnitude of the rate of energy extraction under a set of assumptions. We then move on to the second stage, which uses a viscoelastic tidal model \citep{2013ApJ...767..128C,2014A&A...571A..50C}, in which there is an unknown parameter $\tau$. We calibrate this $\tau$ by varying it until the resulting energy extraction rate matches that obtained in the first stage. The rationale for conducting such a two-step simulation is that the two stages overcome each other's shortcomings: the assumptions made for constructing the equilibrium tidal model are rigid and extreme, including demanding that the inner binary components are of equal mass, and work only for very special hierarchical triple systems, as well as making a host of approximations that may undermine its accuracy regarding the finer details; the viscoelastic model, on the other hand, suffers from no such problems, but has the aforementioned unknown parameter $\tau$, for which there is no established method of calculating through first principles. 

For our following work simulating a set of hierarchical triples under varying parameters, we opt to use the second stage only, leaving $\tau$ as a free parameter in our final empirical fit. This decision is due to the fact that the first-stage simulations previously conducted operates under assumptions that break down for much of the parameter space over which we vary, notably for different values of $m_1$ and $m_2$.

As a starting point for our following investigation, we revert to our hypothetical scenario previously presented in GCEH18. The initial parameters of this system are presented in Table \ref{sim_params}, where $a_1$ and $a_2$ are, respectively, the semimajor axes of the inner and outer orbits, $e_1$ and $e_2$ are the eccentricities of the inner and outer orbits, $i$ is the inclination angle between the two orbits, $m_1$ and $m_2$ are the masses of the inner binary, $m_3$ is the mass of the tertiary, $R_3$ is the radius of the tertiary, and $\tau$ is the viscoelastic relaxation time mentioned above. When simulating this system, we select initial positions such that the vectors for $a_1$ and $a_2$ are perpendicular to each other, and initial velocities for each of the bodies such that all orbits are circular and coplanar. We simulate the system's tidal evolution for $10^5$ years, during which the shrinkage of the inner binary's orbit behaves linearly. 

\begin{table}
	\centering
	\caption{Initial parameters for our simulations, both for the hypothetical scenario which we use as a starting point, as well as the range of values that each parameter was varied over.}
	\label{sim_params}
	\begin{tabular}{ccc}
		\hline
		Parameter & Hypothetical Scenario & Range Varied Over \\
		\hline
                $a_1$/AU & 0.2 & 0.04 - 0.4 \\
                $a_2$/AU & 2.0 & 2.0 - 3.8 \\
                $e_1$ & 0 & - \\
                $e_2$ & 0 & - \\
                $i$ & 0 & - \\
                $m_1$/$M_{\odot}$ & 0.8 & 0.15 - 1.5 \\
                $m_2$/$M_{\odot}$ & 0.8 & see text \\
                $m_3$/$M_{\odot}$ & 1.6 & 0.2 - 1.6 \\
                $R_3$/$R_{\odot}$ & 100 & 20 - 200 \\
                $\tau$/years & 0.534 & 0.1 - 1 \\
		\hline
	\end{tabular}
\end{table}

We then proceed to vary the parameters $m_{\rm 1}$, $q=m_1/m_2$, $m_3$, $R_3$, $\tau$, $a_1$, and $a_2$ one by one, while keeping the other parameters constant, and performing the same simulation. The ranges over which we vary these parameters is also presented in Table \ref{sim_params}. This results in a set of energy extraction rates for different parameters, to which we then perform the desired empirical fit. While not, strictly speaking, a grid of simulations in parameter space, this set of simulations will prove insightful, as demonstrated below.

To speed up our calculations, we translate our original simulation code from 8th-order Runge Kutta to a variable steplength Bulirsch-Stoer algorithm, which costs only a tenth of the calculation time of the original implementation used in GCEH18. We check the consistency of the two codes by repeating our previous simulation runs using our new code, the results of which we find to be practically identical to our original ones.

\subsection{Calculation of Energy Extraction Rate from the Inner Binary}

\begin{figure}
\includegraphics[scale=0.32, angle=270, trim= 0cm 0cm 0cm 0cm]{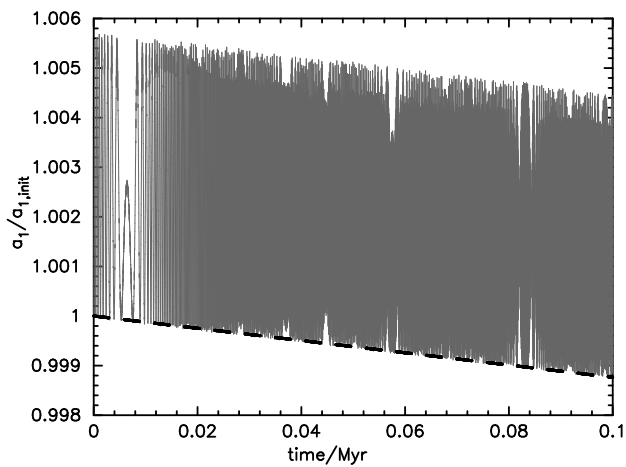}
\caption{Typical example of the oscillatory evolution of the inner binary orbital separation under the influence of tertiary tides. The apparent beating pattern is not a physical phenomenon, but is rather due to Nyquist frequency (aliasing) issues introduced by the plotting process. The dashed line, which traces the positions of the minima of this oscillation, is plotted by connecting the lowest minimum among the first 10 minima, and the lowest minimum among the last 10 minima of this oscillation. In our subsequent plots, we use this method to plot the evolution of these minima values, since the resulting line is aesthetically identical to the actual evolution curves of these minima. \label{osc}}
\end{figure}

As seen in GCEH18, the magnitude of the inner binary orbital separation oscillates as it shrinks, leading to an inner binary orbital energy value that is constantly oscillating as it becomes smaller. A typical example of this is shown in Fig.~\ref{osc}. For the simulations conducted in GCEH18, the amount by which the inner orbit shrinks is so much that this oscillation becomes little more than an insignificant background noise. However, in the simulations below, the inner binary orbital shrinkage is comparable in magnitude or even smaller than this oscillation over the simulated period of $10^5$ years in many instances. Consequently, we cannot simply compare the initial and final inner orbital energies at the beginning and end of the simulation to obtain the energy extraction rate as we did previously. To overcome this oscillation, we take the initial inner binary orbital energy to be that at the first minimum which occurs in our $a_1$, and the final inner binary orbital energy to be that at the last minimum. It is the difference between these two values that we use for the amount of energy extracted. It should be noted here that we also checked the corresponding values calculated using the first and last maxima instead of the minima, as well as the average of the two, and find no significant difference between the results. The fact that we use minima is just an arbitrary preference.

To estimate the precision of our simulations, we set the tertiary radius to progressively smaller values in the hypothetical scenario listed in Table \ref{sim_params}, at some point of which TTs ought to become so insignificant that the apparent amount of energy extracted is dominated by the intrinsic errors of our method. We then take the noise level recovered in this way to be the errors of our simulations. This noise level was found to be $2{\times}10^{33}J$ over a simulation of $10^5$ years.

\section{Results}

In this section, we will present the amount of energy extracted from the inner binary in a set of simulations where the parameters of the triple system are varied. The orbital energy is connected to the inner binary orbital separation by
\begin{equation}
a_{\rm 1}{\propto}-\frac{1}{E_{\rm in}},
\label{Ea}
\end{equation}
\noindent where $a_{\rm 1}$ is the inner binary orbital separation, and $E_{\rm in}$ in the inner binary orbital energy (note that it is always smaller than zero). For constant values of $m_{\rm 1}$, $m_{\rm 2}$, and $m_{\rm 3}$ during the course of evolution, assuming small ${\rm d}a_{\rm 1}$ and ${\rm d}E_{\rm in}$, we have the approximation
\begin{equation}
\frac{{\rm d}a_{\rm 1}}{a_{\rm 1}}=\frac{{\rm d}E_{\rm in}}{E_{\rm in}}.
\label{appEa}
\end{equation}
\noindent Therefore, for the simulation runs mentioned in this paper, we will be presenting the amount of energy extracted from the inner binary in terms of ${\Delta}E/E_{\rm in}$, which is also a proxy for the amount by which the inner binary orbital separation has shrunk.

\subsection{Influence of Inner Binary Masses}

\begin{figure}
\includegraphics[scale=0.32, angle=270, trim= 0cm 0cm 0cm 0cm]{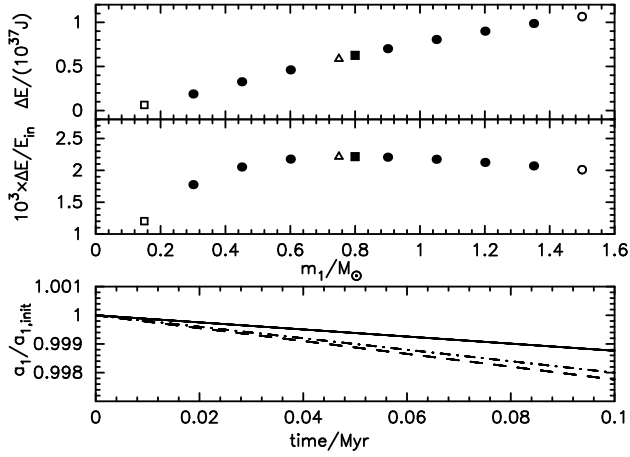}
\caption{Starting with the hypothetical system described in Table \ref{sim_params} (represented by the filled square), $m_{\rm 1}$ is varied while keeping all other parameters constant. Plotted are the results of ten simulation runs, each for $10^5$ years. Plotted in  the upper panel is the amount of energy extracted from the inner binary in each run as a function of $m_{\rm 1}$. Plotted in the upper half of the upper panel is the absolute amount of energy extracted within those $10^5$ years (${\Delta}E$), while the lower half of the upper panel depicts this ${\Delta}E$ in terms of the ratio ${\Delta}E/E_{\rm in}$, where $E_{\rm in}$ is the total orbital energy of the inner binary at the beginning of the simulation (at the beginning of those $10^5$ years). The lower panel plots the evolution of the minima values of each oscillation of the inner binary orbital semimajor axis for some of the models in the upper panel. The solid, dashed, and dash-dotted lines correspond to the models represented by the non-filled square, triangle, and circle in the upper panel, respectively. \label{m1}}
\end{figure}

First of all, we set all the parameters to be equal to those of the hypothetical scenario detailed in Table \ref{sim_params}, with the exception of $m_{\rm 1}$, which we vary from 0.15 to 1.5 M${\odot}$. For each value of $m_{\rm 1}$, we run the simulation for $10^5$ years, during which the inner binary orbital shrinkage is small and behaves linearly, and we expect Eq.\eqref{appEa} to hold. The results of these simulations are presented in Fig.~\ref{m1}, where it can be seen that ${\Delta}E/E_{\rm in}$, and hence the orbital shrinkage, peaks at around $m_{\rm 1}=m_{\rm 2}$. The result of the hypothetical scenario listed in Table \ref{sim_params}, where $m_{\rm 1}$ is actually equal to $m_{\rm 2}$, is also plotted for comparison. We also test the validity of Eq.~\eqref{appEa} by plotting the evolution of the minima of $a_{\rm 1}$ as a function of time, which should provide a straightforward idea of how fast the inner binary orbit actually shrinks. This evolution behaves as expected.

\begin{figure}
\includegraphics[scale=0.32, angle=270, trim= 0cm 0cm 0cm 0cm]{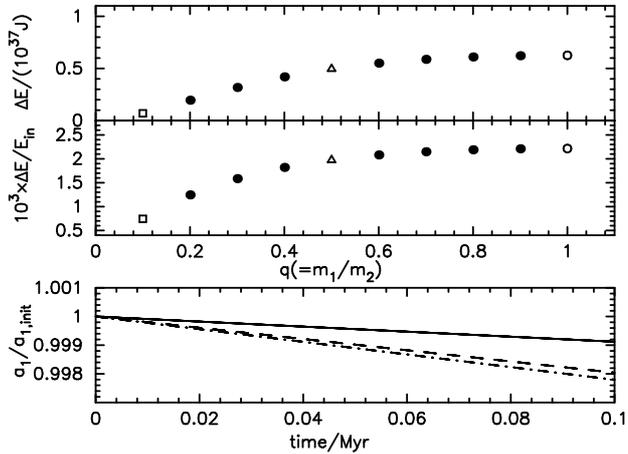}
\caption{Similar to the previous plot, only this time $m_{\rm 1}+m_{\rm 2}$ was kept constant, and the mass ratio $q=m_{\rm 1}/m_{\rm 2}$ is varied. All other parameters are kept constant and equal to the hypothetical system described in Table \ref{sim_params}. The meaning of the panels, symbols, and line styles are the same as those in the previous plot. \label{q}}
\end{figure}

So, what if the total inner binary mass remains constant, but the mass ratio $q=m_{\rm 1}/m_{\rm 2}$ varies? Varying $q$ between 0.1 and 1 while keeping $m_{\rm 1}+m_{\rm 2}$ constant at 1.6M${\odot}$, we arrive at Fig.~\ref{q}, where it can be seen that the energy extraction rate decays slowly as $q$ approaches 0. This is to be expected, as the gravitational potential variation at the tertiary due to the inner binary orbit decreases with smaller $q$.

\begin{figure}
\includegraphics[scale=0.32, angle=270, trim= 0cm 0cm 0cm 0cm]{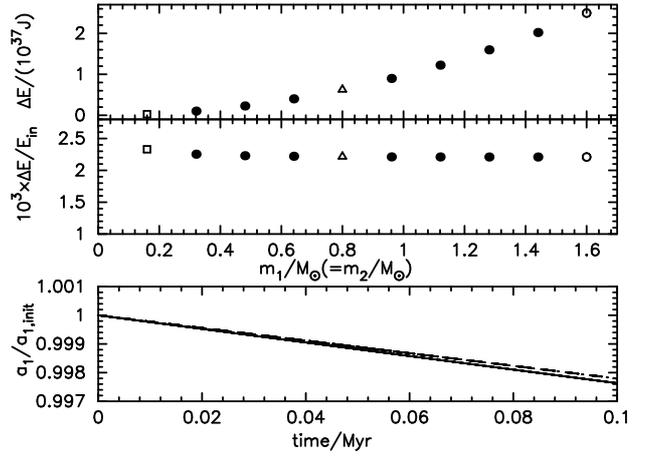}
\caption{In this figure, $q$ is kept constant at 1, while the total value of $m_{\rm 1}+m_{\rm 2}$ is varied. All other parameters are kept constant and equal to the hypothetical system described in Table \ref{sim_params}. The meaning of the panels, symbols, and line styles are the same as those in Fig.~\ref{m1}. \label{m12}}
\end{figure}

Next, we set $q$ to be constant at 1, but vary $m_{\rm 1}$, and the result is plotted in Fig.~\ref{m12}. It appears that ${\Delta}E/E_{\rm in}$ changes little, regardless of what masses are given for the inner binary, as long as $q$=1.

\begin{figure}
\includegraphics[scale=0.32, angle=270, trim= 0cm 0cm 0cm 0cm]{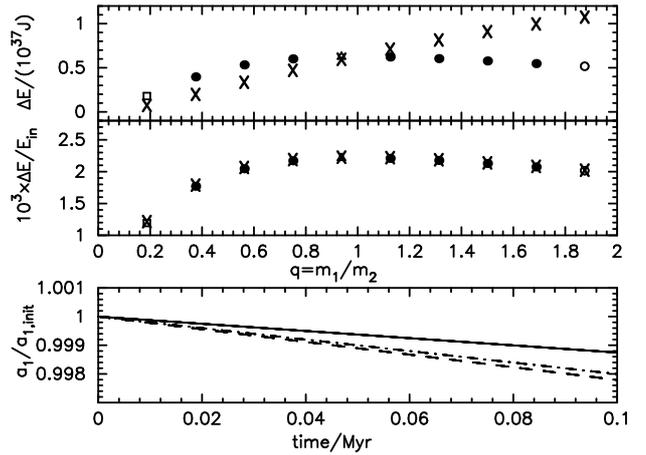}
\caption{Keeping $m_{\rm 1}+m_{\rm 2}$ constant at 1.6 solar masses, $q$ is varied to match the mass ratios induced by varying m1 in Fig.~\ref{m1}. The data points from Fig.~\ref{m1} are plotted over these results as crosses. The meaning of the panels, symbols, and line styles are the same as those in Fig.~\ref{m1}. \label{qalt}}
\end{figure}

Comparing Figs. \ref{m1}, \ref{q}, and \ref{m12}, it is apparent that the maximum value for ${\Delta}E/E_{\rm in}$ is achieved when $q$ is about equal to 1, and that this maximum value does not change much with the values of $m_{\rm 1}$ and $m_{\rm 2}$. So do the values of $m_{\rm 1}$ and $m_{\rm 2}$ matter at all, given that $q$ remains the same? To check this, we again vary $q$ while keeping $m_{\rm 1}+m_{\rm 2}$ constant at 1.6M${\odot}$, but this time we set $q$ to exactly the same values inadvertently obtained in Fig.~\ref{m1}, where we varied $m_{\rm 1}$ while keeping  $m_{\rm 2}$ constant. The results of this set of simulations are displayed in Fig.~\ref{qalt}, plotted over the results shown in Fig.~\ref{m1}. Indeed, it can be seen that ${\Delta}E/E_{\rm in}$ is insensitive to the absolute values of the inner binary masses, as long as $q$ remains the same.

Why is this the case? To answer this question, we refer to the calculations done in the Appendix A of GCEH18, culminating in their Eq. A14, repeated below:

\begin{equation}
{\Delta}E_{\rm P}{\sim}{\frac{135}{4}}\frac{Gm^{2}R_{\rm 3}^{5}a_{\rm 1}^{2}}{a_{\rm 2}^{8}},
\label{deltap}
\end{equation}

\noindent where $E_{\rm P}$ is the greatest self-gravitational potential difference that $m_{\rm 3}$ can undergo, assuming that it is always at equilibrium tide, $G$ is the gravitational constant, and $m=m_{\rm 1}=m_{\rm 2}$ (the equation was derived under the assumption that $m_{\rm 1}=m_{\rm 2}$). While $E_{\rm P}$ has little to do with the actual energy extraction rate (the two are only equivalent if $m_{\rm 3}$ is constantly at equilibrium tide and tidal dissipation is infinitely efficient, neither of which is ever the case), it does decree the absolute upper limit that can be extracted within 1/4 of an inner binary orbit, and thus this expression provides some interesting insights as to how the gravitational field difference at $m_{\rm 3}$ scales with the inner masses. As in the equation, the capability of the inner binary to distort the tertiary scales proportionally with $m^2$, while the energy required to shrink $a_{\rm 1}$ by a certain factor is also proportional to $m^2$. Thus, it shouldn't come as too much of a surprise that ${\Delta}E/E_{\rm in}$ is invariant with $m$.

 \subsection{Influence of the Tertiary}

\begin{figure}
\includegraphics[scale=0.32, angle=270, trim= 0cm 0cm 0cm 0cm]{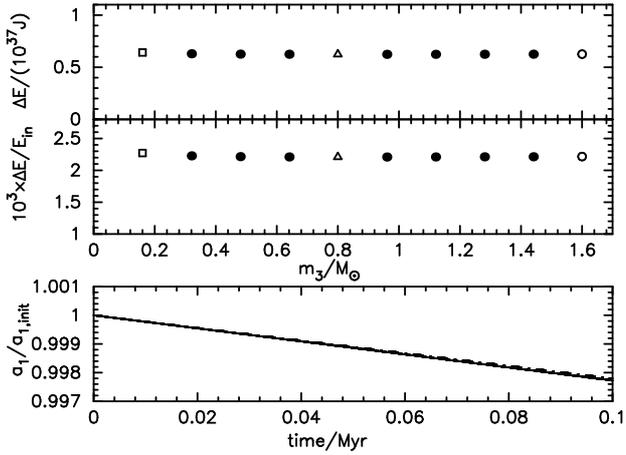}
\caption{In this figure, the tertiary mass $m_{\rm 3}$ is varied, with all other parameters being constant and equal to the hypothetical system described in Table \ref{sim_params}. The panels, symbols, and line styles mean the same as those in Fig.~\ref{m1}. \label{m3}}
\end{figure}

So far, we know that, regarding the masses of the inner binary, only their relative mass ratio is important to ${\Delta}E/E_{\rm in}$. But what about the tertiary? Varying $m_{\rm 3}$ while keeping all other parameters constant, we arrive at Fig.~\ref{m3}, where it can be seen that $m_{\rm 3}$ hardly matters at all. This is probably due to the fact that the amount of energy carried in the tidal bulges on the tertiary is invariant with the mass of the teriary - under the approximation of small tidal bulges, given the same amount of tidal force, the height of the bulges is inversely proportional to the local surface gravitational acceleration ($g$) of the tertiary, whereas the amount of gravitaional potential energy stored per unit height of the bulge is proportional to $g$.

\begin{figure}
\includegraphics[scale=0.32, angle=270, trim= 0cm 0cm 0cm 0cm]{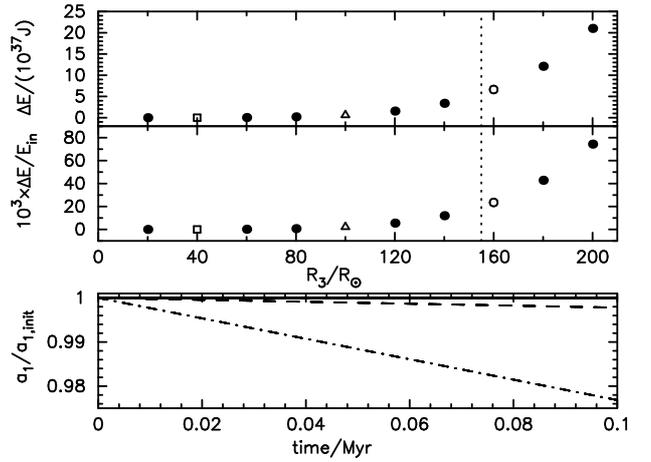}
\caption{Here, the tertiary radius $R_{\rm 3}$, which we know to be a very influential factor in determining orbital shrinkage rate, is varied. The panels, symbols, and line styles mean the same as those in Fig.~\ref{m1}. Note that $m_{\rm 3}$ fills its Roche Lobe in between $R_{\rm 3}$=140 and 160 R${\odot}$ (approximate value indicated by vertical dotted line), so the dash-dotted line is a generous upper limit of how fast TTs can shrink the inner binary orbit in this system without encountering RLOF. \label{R3}}
\end{figure}

If Eq.~\eqref{deltap} is to be believed, ${\Delta}E$ ought to scale proportionately to $R_{\rm 3}^5$ for 100\% dissipation efficiency. However, since we do not know how the tertiary tidal dissipation efficiency scales with $R_{\rm 3}$, it is still worthwhile to calculate how  ${\Delta}E/E_{\rm in}$ evolves with $R_{\rm 3}$. We plot this function for future fitting in Fig.~\ref{R3}, where we see that the influence of $R_{\rm 3}$ is indeed very great.

\begin{figure}
\includegraphics[scale=0.32, angle=270, trim= 0cm 0cm 0cm 0cm]{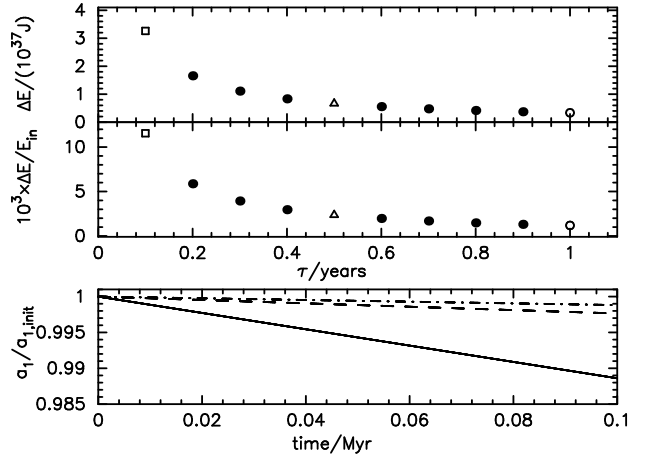}
\caption{Varying the viscoelastic relaxation parameter ${\tau}$, we arrive at this plot. The panels, symbols, and line styles mean the same as those in Fig.~\ref{m1}. \label{tau}}
\end{figure}

Finally, we vary the viscoelastic relaxation parameter ${\tau}$ while keeping the other parameters constant, arriving at Fig.~\ref{tau}.

\subsection{Influence of the Orbital Separations}

\begin{figure}
\includegraphics[scale=0.32, angle=270, trim= 0cm 0cm 0cm 0cm]{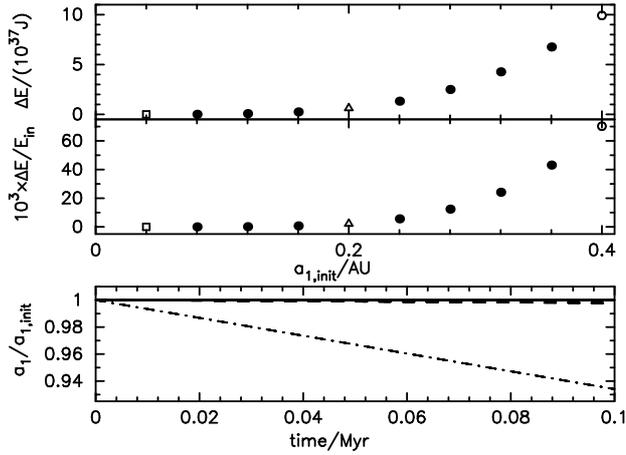}
\caption{Similar to previous plots, only this time the initial value of the inner binary orbital separation $a_{\rm 1}$ is varied. The panels, symbols, and line styles mean the same as those in Fig.~\ref{m1}. \label{a1}}
\end{figure}

\begin{figure}
\includegraphics[scale=0.32, angle=270, trim= 0cm 0cm 0cm 0cm]{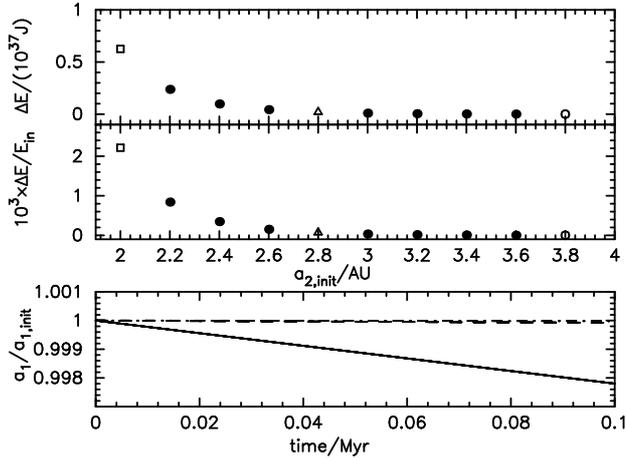}
\caption{The outer binary separation $a_{\rm 2}$ is varied in this figure. The panels, symbols, and line styles mean the same as those in Fig.~\ref{m1}. \label{a2}}
\end{figure}

Regarding the influence of the orbital separations on ${\Delta}E/E_{\rm in}$, one can easily surmise from Eq.~\eqref{deltap} that it is very great. Plotting how  the latter responds to a change in the former in Figs.  \ref{a1} and \ref{a2}, one can see that this is again the case. 

For $a_{\rm 1}$, one can see that, the larger its value the stronger the effects of TTs. This is largely due to the binary quadrupole moment being larger with increasing $a_{\rm 1}$. However, it should be cautioned that too large an $a_{\rm 1}$ can potentially drive the triple system to instability \citep[e.g.][]{2001MNRAS.321..398M,1995ApJ...455..640E}. Conversely, the inner orbital shrinkage stalls once $a_{\rm 1}$ is too small, as once this happens, the change in the gravitational potentialof the inner binary induced by its orbital motion vanishes.

As for $a_{\rm 2}$, one would expect that, as with all tidal phenomena, tidal effects vanish quickly with increasing distance between the body undergoing tidal distortion and the rest of the system.

Throughout the plots displayed up to this point, it should be noted that the error levels are too small to warrant the addition of error bars to these plots. This will be demonstrated in the course of fitting our desired empirical function, which we proceed to do below.

\subsection{Empirical Function Fits}

To obtain the empirical function that is the final goal of this work, we assume that, over $10^5$ years, 

\begin{equation}
\begin{split}
{\Delta}E/E_{\rm in}&=f\left(q,R_{\rm 3},a_{\rm 1},a_{\rm 2},{\tau}\right)\\
&=f_1\left(q\right)f_2\left(R_{\rm 3}\right)f_3\left(a_{\rm 1}\right)f_4\left(a_{\rm 2}\right)f_5\left({\tau}\right).
\end{split}
\label{func}
\end{equation}

\begin{figure}
\includegraphics[scale=0.32, angle=270, trim= 0cm 0cm 0cm 0cm]{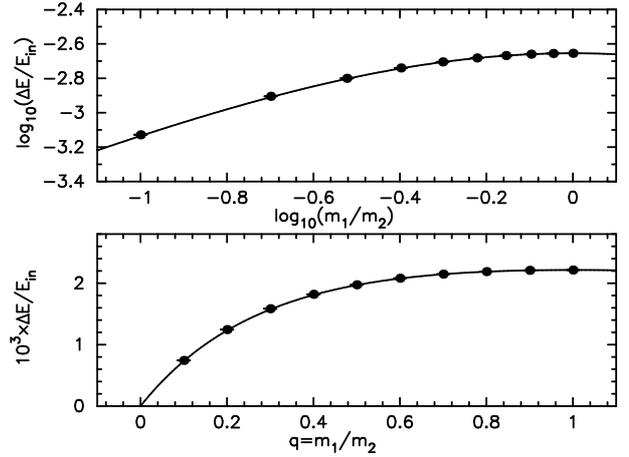}
\caption{Weighted least-squares fit to $f_{\rm 1}=4q/\left(1+q\right)^2$. The upper panel plots the fit in log-log form, while the lower panel plots the fit in linear form. The almost-invisible error bars were calculated assuming a $2{\times}10^{33}$J discrepancy in ${\Delta}E$ both ways. \label{qfit}}
\end{figure}

\noindent Note that, here, we have already taken advantage of our knowledge that $q$ is the only factor by which the masses influence the inner binary orbital shrinkage. After much experimentation, we find that

\begin{equation}
f_{\rm 1}=q\left(\frac{1+q}{2}\right)^{-2}=\frac{4q}{\left(1+q\right)^2}
\label{func1}
\end{equation}

\noindent yields the most sensible fit for $f_{\rm 1}$. This probably has a physical explanation, in that the binary quadrupole moment is the factor that drives TTs. The results of the fit are displayed in Fig.~\ref{qfit}.

\begin{figure}
\includegraphics[scale=0.32, angle=270, trim= 0cm 0cm 0cm 0cm]{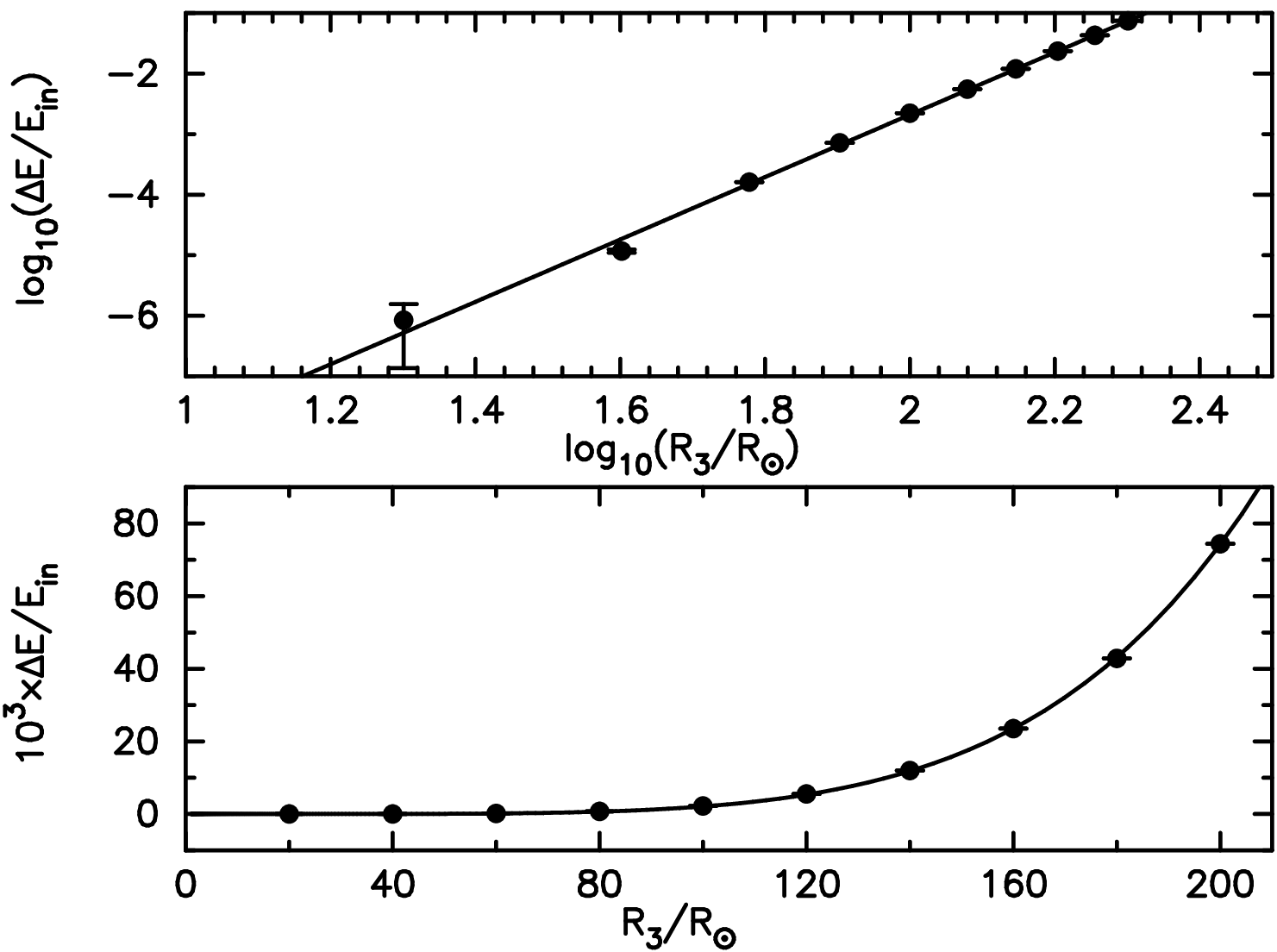}
\caption{Least-squares weighted fit to $f_{\rm 2}=A_{\rm 2}R_{\rm 3}^{C_{\rm 2}}$, with weights applied according to Eq.~\eqref{weight}. It was found that $C_{\rm 2}$=5.2. The upper and lower panels are the same plot in linear space and logarithmic space, respectively. Error bars are plotted for both panels, but are not easily visible for the lower panel. \label{R3fit}}
\end{figure}

\begin{figure}
\includegraphics[scale=0.32, angle=270, trim= 0cm 0cm 0cm 0cm]{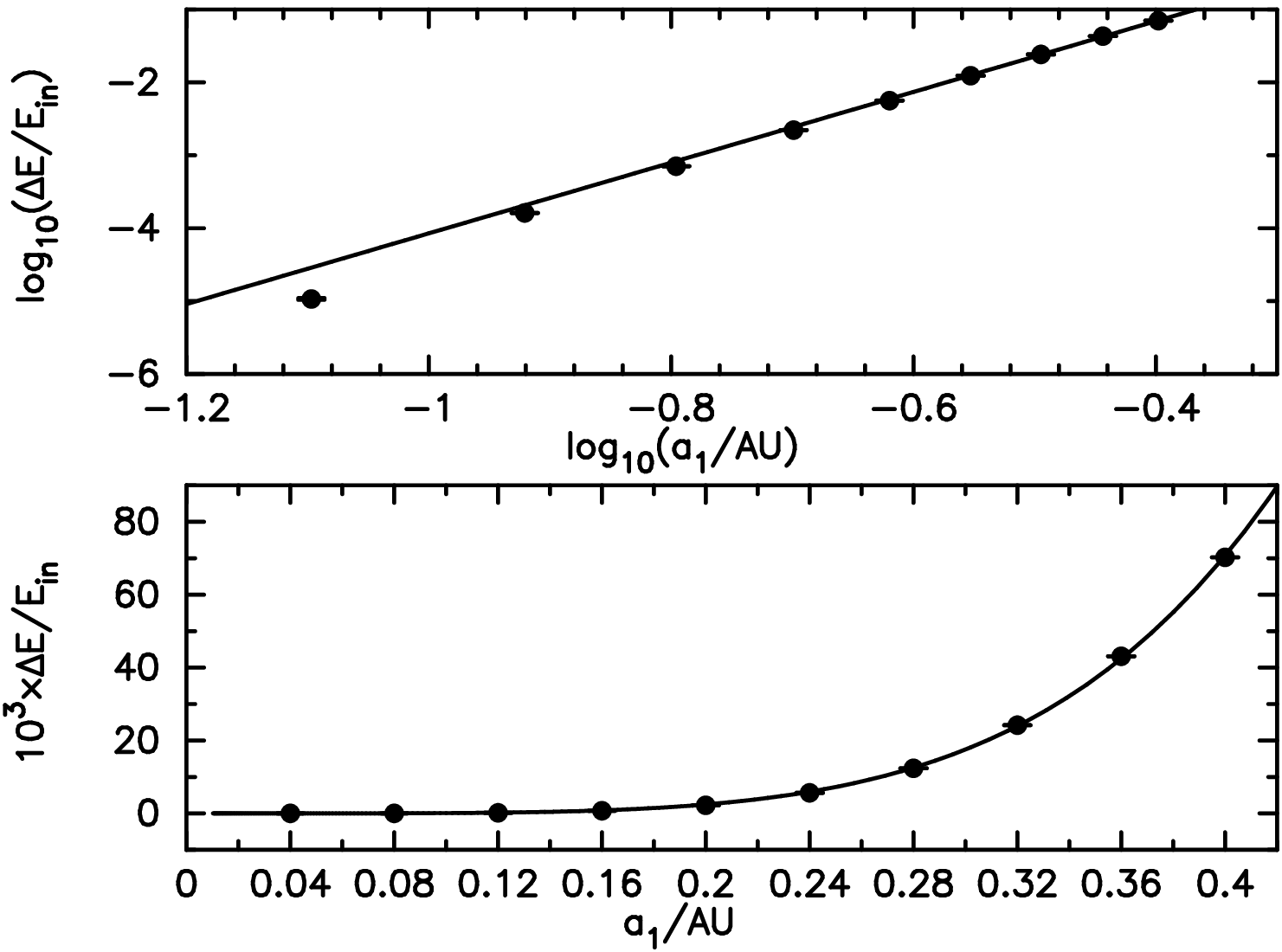}
\caption{Least-squares weighted fit to $f_{\rm 3}=A_{\rm 3}a_{\rm 1}^{C_{\rm 3}}$ with weights applied according to Eq.~\eqref{weight}. It was found that $C_{\rm 3}$=4.8. The upper and lower panels are the same plot in linear space and logarithmic space, respectively. Error bars are plotted for both panels, but are not easily visible in either panel. \label{a1fit}}
\end{figure}

\begin{figure}
\includegraphics[scale=0.32, angle=270, trim= 0cm 0cm 0cm 0cm]{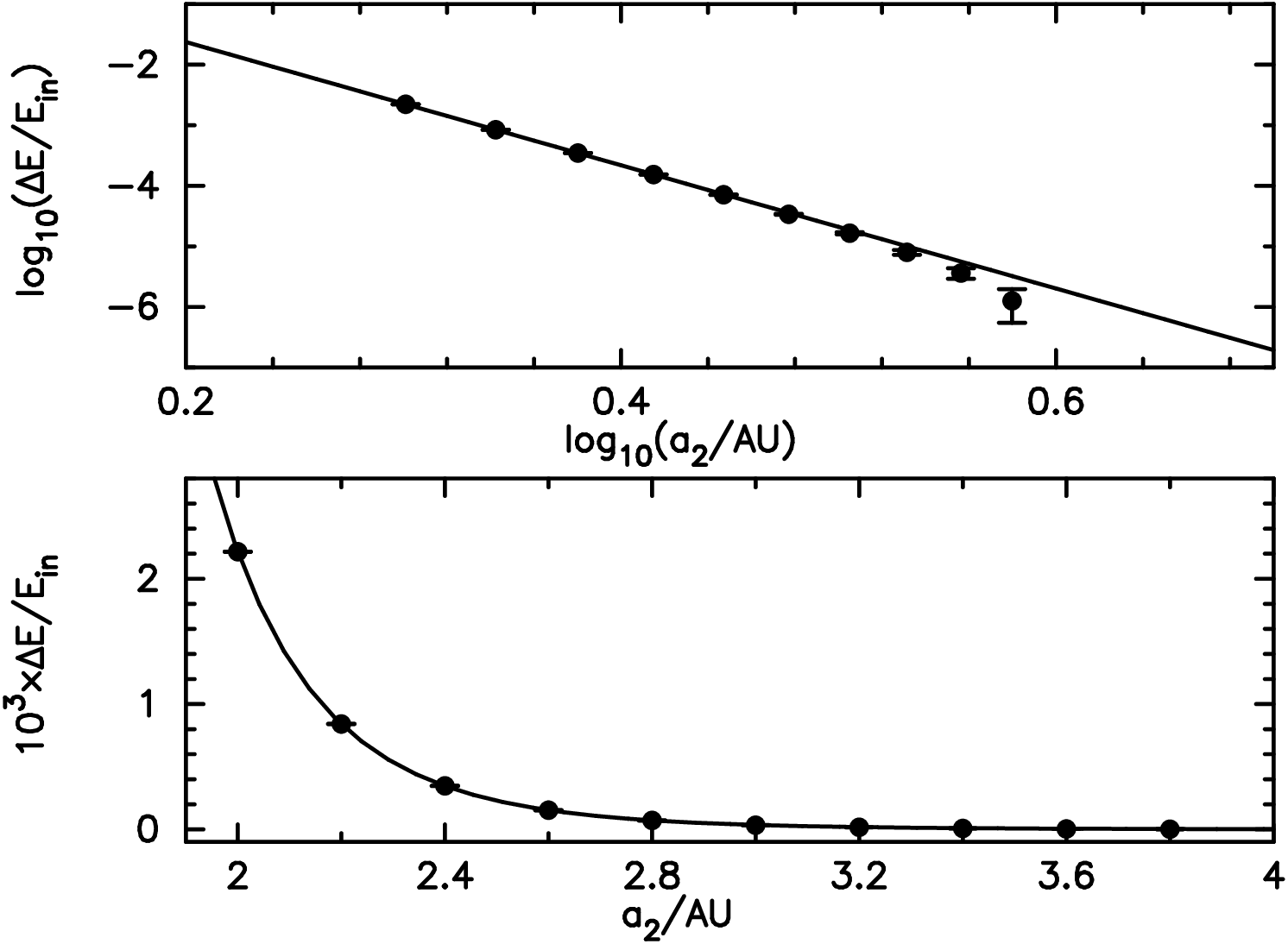}
\caption{Least-squares weighted fit to $f_{\rm 4}=A_{\rm 4}a_{\rm 2}^{C_{\rm 4}}$ with weights applied according to Eq.~\eqref{weight}. It was found that $C_{\rm 4}$=-10.2. The upper and lower panels are the same plot in linear space and logarithmic space, respectively. Error bars are plotted for both panels, but are not easily visible for the lower panel. \label{a2fit}}
\end{figure}

\begin{figure}
\includegraphics[scale=0.32, angle=270, trim= 0cm 0cm 0cm 0cm]{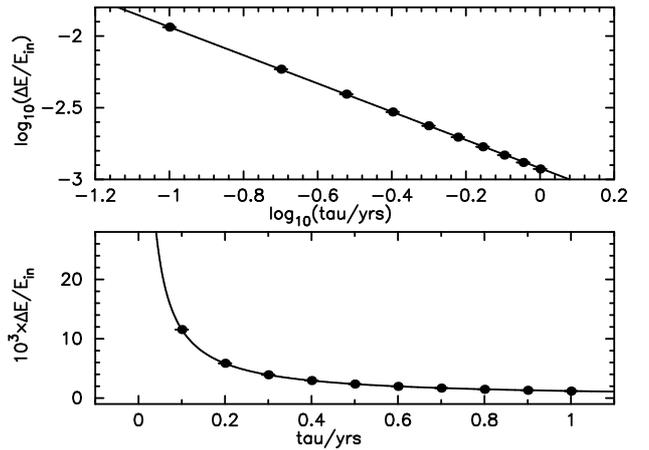}
\caption{Least-squares weighted fit to $f_{\rm 5}=A_{\rm 5}{\tau}^{C_{\rm 5}}$ with weights applied according to Eq.~\eqref{weight}. It was found that $C_{\rm 5}$=-1.0. The upper and lower panels are the same plot in linear space and logarithmic space, respectively. Error bars are plotted for both panels, but are not easily visible in either panel. \label{taufit}}
\end{figure}

As for $f_{\rm 2}$, $f_{\rm 3}$, $f_{\rm 4}$, and $f_{\rm 5}$, we expect these functions to follow some sort of power law, and therefore we plot the data for these fits in logarithmic space, and perform linear fits to these data points using a least-squares algorithm. Attention should be paid here to the fact that our previous error estimates of $2{\times}10^{33}$J, while negligible in linear space, can be significant in logarithmic space, and therefore we calculate the error bars in logarithmic space for each data point, and weight each point by a factor of 

\begin{equation}
w=\left[ \log_{10}\left({\Delta}E+{\sigma}\right) - \log_{10}\left({\Delta}E-{\sigma}\right) \right]^{-2}
\label{weight}
\end{equation}

\noindent when applying the fitting, where $\sigma$ is the error value of $2{\times}10^{33}$J we previously determined. The effect of adding this weighting to our least-squares approach is equivalent to applying a ${\chi}^2$ fit. The optimum power-law indices for $f_{\rm 2}\left(R_{\rm 3}\right)$, $f_{\rm 3}\left(a_{\rm 1}\right)$, $f_{\rm 4}\left(a_{\rm 2}\right)$, and $f_{\rm 5}\left({\tau}\right)$ are found to be 5.2, 4.8,-10.2, and -1.0, respectively, and the fitting functions considered to be optimal are displayed in Figs. \ref{R3fit}, \ref{a1fit}, \ref{a2fit}, and \ref{taufit}. These results are listed in Table \ref{fits}. In general, $R_{\rm 3}$ and $\tau$ yield robust fits, whereas those for $a_{\rm 1}$ and $a_{\rm 2}$ are a little bit more suspicious towards the low energy extraction end, hinting that there is more to the story than a simple power law. However, since low values of ${\Delta}E/E_{\rm in}$ translates into a negligible TT effect, this should not be too much of an issue when applying our fits with a view to simulating the effects of TTs in a stellar evolution algorithm.


\begin{table}
	\centering
	\caption{Results of power-law fits used to find the empirical functions $f_{\rm 2}\left(R_{\rm 3}\right)$, $f_{\rm 3}\left(a_{\rm 1}\right)$, $f_{\rm 4}\left(a_{\rm 2}\right)$, and $f_{\rm 5}\left({\tau}\right)$.}
	\label{fits}
	\begin{tabular}{ccc}
		\hline
		$f_{\rm X}$ & $C_{\rm X}$ \\
		\hline
                $f_{\rm 2}\left(R_{\rm 3}\right)$ & 5.15 \\
                $f_{\rm 3}\left(a_{\rm 1}\right)$ & 4.85 \\
                $f_{\rm 4}\left(a_{\rm 2}\right)$ & -10.17 \\
                $f_{\rm 5}\left({\tau}\right)$ & -0.98 \\
		\hline
	\end{tabular}
\end{table}

It should be noted here that these results invariably deviate from the values found in Eq.~\eqref{deltap}, for a variety of reasons. For $f_{\rm 2}\left(R_{\rm 3}\right)$, it is probably due to the fact that Eq.~\eqref{deltap} was derived under the assumption of small tidal distortions in $m_{\rm 3}$, an assumption which no longer holds when $R_{\rm 3}$ is very large. Also prominent is the issue that ${\tau}$ is very probably intrinsically coupled with the other three in some way, which also explains the discrepancy in the indices for $a_{\rm 1}$ and $a_{\rm 2}$. Lastly, Eq.~\eqref{deltap} was derived for 1/4 of an inner binary orbit under the assumption of 100\% efficient dissipation, while both the orbital period and dissipation efficiency have a degree of dependence on $a_{\rm 1}$, $a_{\rm 2}$ and $R_{\rm 3}$.

Since the nature of how $\tau$ influences the viscoelastic model in our numerical model is exponential, we also attempt an exponential fit for $f_{\rm 5}(\tau)$, in the form of 

\begin{equation}
f_{\rm 5}=C_{\rm 6}\left[   1-e^{-\frac{C_{\rm 7}}{  {\tau}/{\rm yr}    }}    \right].
\label{expotau}
\end{equation}

\noindent which should, in principle, collapse to $f_{\rm 5}=\frac{C_{\rm 6}C_{\rm 7}}{{\tau}/{\rm yr}}$ when ${{\tau}/{\rm yr}}{\gg}C_{\rm 7}$. We find that $C_{\rm 6}=0.198$ and $C_{\rm 7}=0.006$ yields the best fit, and that the fitting curve is identical in appearance to that shown in Fig.~\ref{taufit}. Since we rarely see $\tau$ values as small as 0.006 years, we retain our original power-law fit results for our final empirical function for simplicity, but caution our colleagues that Eq.~\ref{expotau} may be the intrinsically correct formula for $\tau$ dependence for our model. It should also be noted, however, that our model also has a discrepancy relative to the physical world, in the sense that, in the physical world, a $\tau$ value of zero would imply that the tertiary is constantly at equilibrium tide, and no energy is extracted from the inner binary at all. Hence, at extremely small values of $\tau$ much shorter than the steplengths used in our simulations, any orbital shrinkage effects are merely numerical artifacts induced by our algorithm, although exactly at what $\tau$ values this is the case, it is probably extremely hard to tell without extensive real-life observations of triples undergoing TTs.

In summary, we find that ${\Delta}E/E_{\rm in}$ scales proportionally to $4q\left(1+q\right)^{-2}R_{\rm 3}^{5.2}a_{\rm 1}^{4.8}a_{\rm 2}^{-10.2}{\tau}^{-1.0}$ for a fixed simulation time of $10^5$ years. Since ${\Delta}E/E_{\rm in}=2.22{\times}10^{-3}$ for our hypothetical scenario, and noting that the inner binary orbital shrinkage behaves linearly on timescales of $10^5$ years, we can write this as

\begin{equation}
\begin{split}
\frac{ 1  }{  a_{\rm 1}   } \frac{ {\rm d}a_{\rm 1} } { {\rm d}t }=&\left(2.22{\times}10^{-8}{\rm yrs}^{-1}\right)    \frac{4q}{\left(1+q\right)^{2}}\left(\frac{R_{\rm 3}}{100{\rm R}_{\odot}}\right)^{5.2} \\
&\left(\frac{a_{\rm 1}}{0.2{\rm AU}}\right)^{4.8}\left(\frac{a_{\rm 2}}{2{\rm AU}}\right)^{-10.2}\left(\frac{\tau}{0.534{\rm yrs}}\right)^{-1.0}.
\label{result} 
\end{split}
\end{equation}

\section{Discussion}

Before we rush to the conclusion that our empirical function (1/$a_{\rm 1}$)(d$a_{\rm 1}$/d$t$) = $\left(2.22{\times}10^{-8}{\rm yrs}^{-1}\right)$ $4q\left(1+q\right)^{-2}$ $(R_{\rm 3}/100{\rm R}_{\odot})^{5.2}$ $(a_{\rm 1}/0.2{\rm AU})^{4.8}$ $(a_{\rm 2}/2{\rm AU})^{-10.2}$ $({\tau}/0.534{\rm yrs})^{-1.0}$ can approximate the effect of TT energy extraction rates, we should note that our simulations do not account for any correlation terms between the parameters varied. In other words, varying two different parameters at the same time may result in behaviour that deviates from what we expect when varying them one at a time. Also, it should be noted that extrapolating to parameter regimes beyond what has been simulated may also be questionable, since our fits already hint at a deviation from the power laws that we use to fit our results as we go to such regimes. For these reasons, it would be wise to conduct a few tests by comparing extrapolations of this empirical function with simulations of triple systems with parameters different to those simulated throughout the course of this work.

HD181068 ($a_1$=4.777R${\odot}$, $a_2$=90.31R${\odot}$, $e_1$=$e_2$=$i$=0, $m_1$=0.870M${\odot}$, $m_2$=0.915M${\odot}$, $m_3$=3.0M${\odot}$, $R_3$=12.46M${\odot}$, $\tau$=0.88 years) is a triple system which can be used to perform such a test, since some of its parameters lie beyond the range of our simulations in this paper. For HD181068, we find that ${\Delta}E/E_{\rm in}=5.72{\times}10^{-6}$ over $10^5$ years using our empirical function, while a full viscoelastic simulation yields ${\Delta}E/E_{\rm in}=3.31{\times}10^{-6}$ over $10^5$ years. Thus, we can see that there is a significant deviation, but that the function is still accurate to within an order of magnitude.

To check for correlation terms between the parameters, we conduct a set of simulations, the details of which are listed in Table~\ref{testresults}. All the parameters not displayed in the table are identical to those of the hypothetical scenario in Table \ref{sim_params}. As we can see from the simulation results, our empirical function performs admirably in predicting ${\Delta}E/E_{\rm in}$ for these test runs.


\begin{table*}
 \caption{Comparison between empirical function fits and corresponding results of full simulations for $10^5$ years for a set of arbitrarily selected test runs. It can be seen that the errors are relatively small, and that the performance of our empirical function is satisfactory.}
 \label{testresults}
 \begin{tabular}{lllllllllll}
  \hline
    Test \# & $m_{\rm 1}$/M$\odot$ & $m_{\rm 2}$/M$\odot$ & $m_{\rm 3}$/M$\odot$ & $q(=m_{\rm 1}/m_{\rm 2}$) & $a_{\rm 1}$/AU & $a_{\rm 2}$/AU & $R_{\rm 3}$/R$\odot$ & $\tau$/years & Simulated ${\Delta}E/E_{\rm in}$ & Function ${\Delta}E/E_{\rm in}$ \\
    \hline
    1 & 0.12 & 1.21 & 1.7 & 0.099 & 0.16 & 2.1 & 84 & 0.4 & $7.688{\times}10^{-5}$ & $8.170{\times}10^{-5}$ \\
    2 & 0.24 & 1.22 & 1.8 & 0.197 & 0.17 & 2.2 & 88 & 0.5 & $1.116{\times}10^{-4}$ & $1.160{\times}10^{-4}$ \\
    3 & 0.36 & 1.23 & 1.9 & 0.293 & 0.18 & 2.3 & 92 & 0.6 & $1.264{\times}10^{-4}$ & $1.298{\times}10^{-4}$ \\
    4 & 0.48 & 1.24 & 2.0 & 0.387 & 0.19 & 2.4 & 96 & 0.7 & $1.316{\times}10^{-4}$ & $1.340{\times}10^{-4}$ \\
    5 & 0.60 & 1.25 & 2.1 & 0.480 & 0.20 & 2.5 & 100 & 0.8 & $1.318{\times}10^{-4}$ & $1.331{\times}10^{-4}$ \\
    6 & 0.72 & 1.26 & 2.2 & 0.571 & 0.21 & 2.6 & 104 & 0.9 & $1.300{\times}10^{-4}$ & $1.298{\times}10^{-4}$ \\
    7 & 0.84 & 1.27 & 2.3 & 0.661 & 0.22 & 2.7 & 108 & 1.0 & $1.271{\times}10^{-4}$ & $1.252{\times}10^{-4}$ \\
    8 & 0.96 & 1.28 & 2.4 & 0.750 & 0.23 & 2.8 & 112 & 1.1 & $1.219{\times}10^{-4}$ & $1.201{\times}10^{-4}$ \\
    9 & 1.08 & 1.29 & 2.5 & 0.837 & 0.24 & 2.9 & 116 & 1.2 & $1.175{\times}10^{-4}$ & $1.147{\times}10^{-4}$ \\
    10 & 1.20 & 1.30 & 2.6 & 0.923 & 0.25 & 3.0 & 120 & 1.3 & $1.134{\times}10^{-4}$ & $1.094{\times}10^{-4}$ \\
  \hline
 \end{tabular}
\end{table*}

As a further test of our empirical function, we set ($a_{\rm 1}$/$a_{\rm 2}$) to a fixed value of 0.1, and then vary $a_{\rm 1}$ to see how ${\Delta}E/E_{\rm in}$ evolves with changing $a_{\rm 1}$. If the our empirical function were a physical law, then since 
\begin{equation}
{\Delta}E/E_{\rm in}{\propto}a_{\rm 1}^{4.8}a_{\rm 2}^{-10.2},
\end{equation}
\noindent one would expect that, with ($a_{\rm 1}$/$a_{\rm 2}$) being a constant, ${\Delta}E/E_{\rm in}$ should scale as $a_{\rm 1}$ to the power of -5.4. The power-law index that results from our fits is -5.0. While not fatal to the validity of our empirical function, this discrepancy implies that there is more to this relation than meets the eye. Coincidentally, an index of -5.0 is the exact power law one would expect if one were to equate ${\Delta}E$ with the right-hand side of Eq.~\eqref{deltap}, and substitute $E_{\rm in}$ using Eq.~\eqref{Ea}. However, since the relation between ${\Delta}E$ and the right-hand side of Eq.~\eqref{deltap} is a complicated one, we have no reason to believe that a power-law of -5.0 is intrinsic to this relation. Perhaps future studies that improve our understanding of the correlation between $\tau$ and the other orbital parameters may shed more light on this issue.

It should also be drawn to the attention of the reader that our model also suffers from the uncertainty that the second Love number, $k_{\rm 2}$, is difficult to determine for most stars. In our simulations, we used $k_{\rm 2}=0.2$, which is what GCEH18 used, as prescribed by \citealt{2017MNRAS.472.4965Y} for red giants, but actual values of $k_{\rm 2}$ can vary (see, for instance, the value for a real red giant in \citealt{2013MNRAS.428.1656B}). However, since tidal effects, including ${\Delta}E/E_{\rm in}$, scale proportionally with $k_{\rm 2}$, we do not expect this to be too great an issue.

Given our results, how reliably can we expect to incorporate the effect of TTs into stellar evolution codes? As of yet, $\tau$ is still a free parameter in our empirical function fits, subject to further studies which should be aimed at breaking the degeneracy between this parameter and the other orbital parameters. Before these studies are conducted, we expect work that makes use of our current implementation of the fitting function to have an element of arbitrariness to it. However, given the possibility of calibrating this free parameter using other means such as observations, even the current form of our fitting function may have its uses. Of greater inconvenience is the fact that our empirical fitting function can only, as of yet, deal with coplanar, circular orbits. However, studies of TTs in non-coplanar, non-circular orbits would entail disentanglement from other effects present in hierarchical triples, such as Lidov-Kozai resonance  \citep[e.g.][]{2016ARA&A..54..441N}, an issue which is not expected to be resolved anytime soon.

\section*{Acknowledgements}

We thank our colleagues, including but not limited to Alexandre Correia, Zhengwei Liu, and Rosanne Di Stefano for valuable discussion in the course of this work.

This work was jointly supported by the Natural Science Foundation of China (Grant No. 11521303), and the Science and Technology Innovation Talent Programme of Yunnan Province (Grant No. 2017HC018).










\bsp	
\label{lastpage}
\end{document}